\documentclass[prb,twocolumn,superscriptaddress,longbibliography,notitlepage]{revtex4-1}
\usepackage{amssymb}
\usepackage{amsmath}
\usepackage{graphicx}
\usepackage{hyperref}
\usepackage{xcolor}
\usepackage{multirow}
\usepackage{amsfonts}
\usepackage[title]{appendix}

\newcommand{\comment}[1]{}

\begin{document}

\title{Probing $d$-wave pairing in the $t$-$t'$-$U$ Hubbard model with tensor-backflow wave functions}

\author{Xiao Liang}
\email{xliang06@wm.edu}
\affiliation{Research Computing(Information Technology), William \& Mary, Williamsburg, Virginia 23185, USA}
\affiliation{Department of Physics, William \& Mary, Williamsburg, Virginia 23185, USA}

\begin{abstract}

	Building on the recently developed tensor-backflow method, which accurately describes energies and stripe states in the Fermi-Hubbard model, we investigate $d$-wave pairing in the two-dimensional $t$-$t'$-$U$ Hubbard model.
	For the $8\times8$ lattice, we improve the wave function using symmetry projections followed by one Lanczos step, whereas for the $12\times 12$ and $16\times 16$ lattices, we apply one Lanczos step without symmetry projection.
	We focus primarily on the parameter regime with filling $n=0.875$, on-site repulsion $U=8$ and next-nearest-neighbor hopping $t'=-0.2t$, a regime in which enhanced $d$-wave pairing has been reported.
	The resulting energies are competitive with those obtained using state-of-the-art neural quantum states. For example, on a $16\times 16$ lattice with periodic boundary conditions, the tensor-backflow wave function after one Lanczos step without explicit symmetry enforcement achieves a relative energy difference of $4.4\times 10^{-3}$ from the symmetry-preserving neural quantum state result.
	From the real space pair correlations, for $t'=-0.2t$, we observe stronger $d$-wave pair correlations together with the characteristic relative sign between horizontal and vertical bond-pair components.
	Compared with $t'=0$, finite negative $t'$ increases the spectral weight of the corresponding pair-density matrix, which indicates the enhanced overall $d$-wave pair fluctuations.
	The leading eigenvalue is enhanced on the $8\times8$ lattice for both with and without symmetry projections.
	For larger lattices such as $12\times 12$ and $16\times 16$, the leading pair-density matrix spectral weight is distributed among several nearly degenerate eigenmodes rather than being concentrated in a single dominant mode. This indicates enhanced and more broadly distributed $d$-wave pair fluctuations, but no single eigenvalue exhibits the extensive scaling required for ODLRO.

\end{abstract}
\maketitle

\section{Introduction}
\label{sec:introduction}
The Fermi-Hubbard model is a paradigmatic model for studying the microscopic origin of unconventional superconductivity~\cite{Hubbard_SC1,Hubbard_SC2}.
Determining its ground state is challenging because several competing phases can be extremely close in energy~\cite{Hubbard_solve1,Hubbard_solve2,Hubbard_solve3,Hubbard_solve4,Hubbard_solve5,Hubbard_solve6,Hubbard_solve7,Hubbard_solve8}.
On finite lattices, the ground state can also depend sensitively on system size and boundary conditions.
Therefore, to reduce the finite-size effect, the lattice size should be sufficiently large.
The $t$-$t'$-$U$ Hubbard model is more challenging than the pure Hubbard model.
The next-nearest-neighbor hopping $t'$ changes the Fermi surface and makes the ground state more sensitive to finite-size effects. Meanwhile, several nearly degenerate phases compete near the ground state.

Variational Monte Carlo (VMC) provides a flexible framework for approximating many-body ground states and their energies.
Unlike many projector quantum Monte Carlo approaches, VMC does not suffer from a fermionic sign problem in sampling the probability distribution of a prescribed variational wave function~\cite{sign_problem_1,sign_problem_2}; its accuracy instead depends on the expressiveness and optimization of the ansatz.
Neural quantum states (NQSs) provide a powerful class of variational wave functions. Neural-network wave functions have achieved high accuracy for a variety of strongly correlated systems, including many-electron problems in continuous space~\cite{VMC_NN_continue_1,VMC_NN_continue_2,VMC_NN_continue_3,VMC_NN_continue_4}.

For lattice models, an effective variational wave function must combine sufficient expressive power with tractable optimization.
In VMC, the variational parameters are typically optimized using gradient-based methods to minimize the energy expectation value.
However, the effectiveness of VMC depends crucially on the quality of the variational wave function.
Incorporating prior physical information, such as known sign structures or Hamiltonian symmetries, can substantially improve optimization and variational accuracy.
The frustrated $J1$-$J2$ model provides one example.
Near the maximal frustration $J2=0.5J1$, employing a prior Marshall sign rule~\cite{MSR} significantly reduces the optimization difficulty~\cite{J1J2_MSR_1,J1J2_MSR_2,J1J2_MSR_3}.
In addition to incorporating prior sign structure, enforcing Hamiltonian symmetries can reduce the optimization difficulty and improve variational accuracy\cite{symmetry_J1J2}.
This strategy has also proved effective for the Fermi-Hubbard model~\cite{NQS_Pfaffian,symmetry_Hubbard1,symmetry_Hubbard2}.

Recently, the tensor-backflow wave function has shown high efficiency and accuracy in solving the Hubbard model~\cite{TBF}.
For example, combined with one Lanczos step, the tensor-backflow wave function achieves energies competitive with Transformer-based NQS calculations~\cite{Transformer NQS}, meanwhile the method has obtained charge-density-wave (CDW) and spin-density-wave (SDW) with periods of $1/n$ and $2/n$, on the $16\times 16$ lattice with periodic boundary conditions (PBC), where $n$ is the electron filling. As a variational wave function, the tensor-backflow is optimized by first-order-gradient method with sufficient MC sample number, and it has sufficient expressive power for a wide range of parameters in the Fermi-Hubbard model.
For example, for the case of $n=0.875$, $U=8$, $t'=-0.2t$ on the $12\times 12$ under PBC, the tensor-backflow has successfully obtained the width-4 stripe with the energy lower than the Transformer NQS.

In this work, we apply the tensor-backflow method on cases of $n=0.875$, $U=8$, $t'=-0.2t$ on square lattices up to the size $16\times 16$ under PBC and open boundary conditions (OBC).
For the $8\times 8$ case, we perform symmetry projections on the tensor-backflow wave function, followed by a Lanczos step. The best energy achieved in this work is -0.7441, which is $8.1\times 10^{-4}$ and $8.7\times 10^{-3}$ higher than those achieved by the Transformer NQS: -0.7447~\cite{Transformer NQS}, and the fully symmetrized Pfaffian: -0.7506~\cite{symmetry_Hubbard1}, respectively.
For cases of $12\times 12$ and $16\times 16$, a Lanczos step is applied on the wave function.
For the case of $16\times 16$ under PBC, we compare energies from different setups of pinning fields, and we find that the width-4 stripe gives the lowest energy.
The best energy achieved in this work for the $16\times 16$ lattice is -0.7408, this is $6.5\times 10^{-3}$ lower than the Neural Pfaffian~\cite{NQS_Pfaffian} and the relative energy differences with respect to state-of-the-art NQS results~\cite{symmetry_Hubbard2,NQS byteadance} are below $4.5\times 10^{-3}$.
Furthermore, we calculate pair correlations and investigate components of the pair correlation, for both $t'=-0.2t$ and $t'=0$ on all lattice sizes.
For $t'=-0.2t$, we find stronger $d$-wave pair correlations than for $t'=0$, together with the characteristic relative sign between horizontal and vertical bond-pair components, such benefit of $t'$ is also demonstrated by comparing eigenvalues of pair-density matrices. For the same lattice size, eigenvalues of $t'=-0.2t$ is generally higher than the counterparts of $t'=0$.
However, the largest eigenvalue of $t'=-0.2t$ is not higher than that of $t'=0$ for lattices of $12\times 12$ and $16\times 16$.

\section{Tensor-Backflow method}
\label{sec:tensor-backflow method}

We briefly review the tensor-backflow method introduced previously~\cite{TBF,TBF0}.
The tensor-backflow wave function is constructed based on backflow corrections.
In backflow corrections, the particle position is transformed as~\cite{backflow1}:
\begin{equation}
	\mathbf{r}_{\alpha}^B=\mathbf{r}_{\alpha}+\sum_{\beta}\eta_{\alpha\beta}[\mathbf{S}](\mathbf{r}_\beta-\mathbf{r}_\alpha),
	\label{eq:r_transformation}
\end{equation}
where $\mathbf{r}_\alpha$ are actual particle positions and $\eta_{\alpha\beta}[\mathbf{S}]$ are variational parameters depending on the many-body state $|\mathbf{S}\rangle$. Consequently, the backflow-corrected orbital of a particle on $\mathbf{r}_\alpha$: $\phi_{k}^B(\mathbf{r}_\alpha)$ is a linear combination of orbitals from backflow sites $\mathbf{r}_\delta$~\cite{backflow1,backflow2}:
\begin{equation}
\phi_k^B(\mathbf{r}_\alpha)=\sum_{\mathbf{r}_\delta} \eta_{\alpha\delta}[\mathbf{S}]\phi_k(\mathbf{r}_\delta),
\label{eq:backflow_orbital}
\end{equation}
where $\mathbf{r}_\delta$ denote the sites considered in backflow corrections. For example, for nearest-neighbor (NN) backflow, $\mathbf{r}_\delta$ contains $\mathbf{r}_\alpha$ and the NN sites. For all-site backflow, $\mathbf{r}_\delta$ covers all lattice sites.
In original backflow corrections, for simplicity, $\eta$ depends on $\mathbf{s}(\mathbf{r}_\alpha)$ and $\mathbf{s}(\mathbf{r}_\delta)$ instead of the full many-body state $|\mathbf{S}\rangle$~\cite{backflow1,backflow2}.

A full-rank tensor can represent an arbitrary function of several independent variables.
For the function of $\eta_{\alpha\delta}\phi_k$ in the backflow correction, the independent variables are the particle position $\mathbf{r}_\alpha$, orbital index $k$, configurations of $\mathbf{s}(\mathbf{r}_\alpha)$ and $\mathbf{s}(\mathbf{r}_\delta)$. A tensor built on these variables has the representation ability beyond the product of $\eta_{\alpha\delta}$ and $\phi_k$.
To further improve the representation ability on backflow corrections, we consider the summation index of backflow terms as an additional independent degree of freedom.
Thus, the single tensor is built as: $g_{\mathbf{r}_\alpha,k,\delta,\mathbf{s}(\mathbf{r}_\alpha),\mathbf{s}(\mathbf{r}_\delta)}$~\cite{TBF}.
The resulting wave function is represented by the Slater determinant:
\begin{equation}
w(\mathbf{S})=\mathrm{det}M^B,
	\label{eq:ws}
\end{equation}
where the matrix element is built according to the backflow summation: $M_{i,k}^B=\sum_j g_{\mathbf{r}_i,k,j,\mathbf{s}(\mathbf{r}_i),\mathbf{s}(\mathbf{r}_j)}$, where $i$ and $j$ are used for labeling particles and backflow terms, respectively.

A single full-rank tensor contains a number of variational parameters equal to the product of its dimensions.
For a particular particle filling $n$, the parameter number of $g$ scales quadratically and cubically with lattice size for a finite range backflow and all-site backflow, respectively.
However, only a subset of these parameters contributes to a given amplitude $w(\mathbf{S})$.
The computational complexity for generating one $w(\mathbf{S})$ is $\mathcal{O}(QN^3)$, where $Q$ is the number of backflow terms considered and $N$ is the particle number. For a lattice with $M$ sites, there are $Q$=5, 9 and $M$ for NN, next-nearest-neighbor (NNN) and all-site backflow, respectively.

In the conventional Hartree-Fock construction in Eq.(\ref{eq:backflow_orbital}), the particle and orbital spin labels must match, resulting in a block-diagonal structure of $M^B$ in spin space. This restriction can limit the expressive power for systems with nontrivial spin correlations. As the tensor representation is beyond the original backflow correction built from Hartree-Fock orbitals, all matrix elements in $M_{i,k}^B$ are considered regardless of spins of orbital $k$ and particle $i$.

A Hartree-Fock orbital without spin can be represented by a tensor containing dimensions of $\mathbf{r}_i$ and $k$. Without restricting the spin identity between the orbital and the particle, the unrestricted-Hartree-Fock (UHF) is represented by a tensor with the dimension of $g_{\mathbf{r}_i,k,\sigma_i}^{\mathrm{UHF}}$, where $\sigma_i$ denotes the particle spin.
By expanding the UHF's tensor, the remaining dimensions in the tensor-backflow are added as: $g_{\mathbf{r}_i,k,\sigma_i,2,j,\mathbf{s}(\mathbf{r}_j)}$, where the dimension of 2 is for distinguishing double occupation on $\mathbf{r}_i$.
Because the optimization cost of a UHF is low, the tensor-backflow wave function is initialized by a UHF to reduce the optimization cost.

The tensor representation is optimized by first-order gradient descent method, a parameter in the tensor is updated as~\cite{TBF,SGD_sign1,SGD_sign2}:
\begin{equation}
	x_{t+\Delta t}=x_{t}-\mathrm{sgn}(G_x)d_x,
	\label{eq:SGD}
\end{equation}
where $G_x$ is the gradient of parameter $x$, $d_x$ is the step size.
The energy and gradients are estimated by Markov-Chain-Monte-Carlo process:
\begin{equation}
G_x=2\langle E(\mathbf{S})O_x(\mathbf{S})\rangle-2\langle E(\mathbf{S})\rangle\langle O_x(\mathbf{S})\rangle,
\label{eq:EsOs}
\end{equation}
where local estimators under a configuration $|\mathbf{S}\rangle$ are: $E(\mathbf{S})=\sum_{\mathbf{S}'}w(\mathbf{S}')/w(\mathbf{S})\langle \mathbf{S}'\hat{H}\mathbf{S}\rangle$ and $O_x(\mathbf{S})=\partial \mathrm{ln}|w(\mathbf{S})|/\partial x$.

To improve the energy accuracy of tensor-backflow, subsequent methods such as symmetry projections and a Lanczos step are applied on the wave function. The wave function for symmetry projections is constructed as~\cite{NQS_Pfaffian,symmetry_Hubbard1}:
\begin{equation}
	w^{\mathrm{sym}}(\mathbf{S})=\sum_{\hat{T}\in G} \lambda_G \xi(\hat{T},\mathbf{S}) w(\hat{T}\mathbf{S}),
	\label{qs:ws_sym}
\end{equation}
where $\hat{T}$ applies a symmetry operation on $|\mathbf{S}\rangle$, $\xi(\hat{T}, \mathbf{S})$ is the permutation sign from the symmetry operation and $\lambda_G$ is the character of the symmetry representation.
Another method to improve the wave function is a Lanczos step. It improves the wave function by constructing a state in Krylov subspace~\cite{lanczos}:
\begin{equation}
	|\Psi_{p+1}\rangle=A|\Psi_p\rangle+B|\Psi_p^{\perp}\rangle,
	\label{eq:ws_p1}
\end{equation}
where $|\Psi_p^{\perp}\rangle=(\hat{H}-E)/\sigma|\Psi_p\rangle$, with $\sigma^2$ the energy variance of $|\Psi_p\rangle$.

Both symmetry projections and a Lanczos step lead to multiple Slater determinants in the final wave function.
However, even with the correct kinds of symmetries, symmetry projection can improve energy accuracy when the wave function isn't trapped at local minima.
Meanwhile, a Lanczos step preserves the wave function's symmetry and it can improve the energy accuracy systematically~\cite{J1J2_MSR_2,TBF}.
\begin{table*}[]
\begin{tabular}{ccccccccc}
\hline
\ \ $U$\ \  & \ \ \ \ \ \ $E_{p=1}$\ \ \ \ \ \  & \ \ \ \  & \ \ \ \ \ \ (0,\ 0)\ \ \ \ \ \  & \ \ \ \ \ \ (1,\ 0)\ \ \ \ \ \  & \ \ \ \ \ \ (2,\ 0)\ \ \ \ \ \  & \ \ \ \ \ \ (1,\ 1)\ \ \ \ \ \  & \ \ \ \ \ \ (2,\ 1)\ \ \ \ \ \  & \ \ \ \ \ \ (2,\ 2)\ \ \ \ \ \  \\ \hline
2   &   -1.3360(2)  & this work & 2.068(6) & 0.092(3) & -0.112(4) & 0.114(4) & 0.027(2) & 0.178(3) \\
    &           & exact     & 2.06693 & 0.09223 & -0.11187 & 0.11381 & 0.02840 & 0.17793 \\
4   &   -1.2236(3)  & this work & 2.065(6) & 0.086(3) & -0.093(4) & 0.100(4) & 0.024(3) & 0.153(3) \\
    &           & exact     & 2.06345 & 0.08714 & -0.09422 & 0.10013 & 0.02453 & 0.15302 \\ \hline
\end{tabular}
	\caption{Comparisons on pair correlations obtained from this work and exact results~\cite{pair_correlation_2D_Hubbard}, on the $4\times 4$ lattice with $N_\uparrow=N_\downarrow=5$. The pair separation is $\mathbf{d}=(\mathrm{dx},\mathrm{dy})$. Each estimation is obtained with the $p$=1 wave function and $1\times 10^7$ MC samples. Estimation errors are indicated by the last digit in the parentheses.}
	\label{tab:pair_compare}
\end{table*}

%In backflow corrections, configurations of $\mathrm{s}(\mathrm{r}_i)$ and $\mathrm{s}(\mathrm{r}_j)$ are correlated in the orbital $\phi_k^B$. The tensor representation contains high order correlations of $\mathrm{s}(\mathrm{r}_i)$ and $\mathrm{s}(\mathrm{r}_j)$ naturally.
%Considering more backflow terms includes more correlations into the orbital $\phi_k^B$.

\section{Numerical Results}
\begin{figure}[t]
\includegraphics[width=\columnwidth]{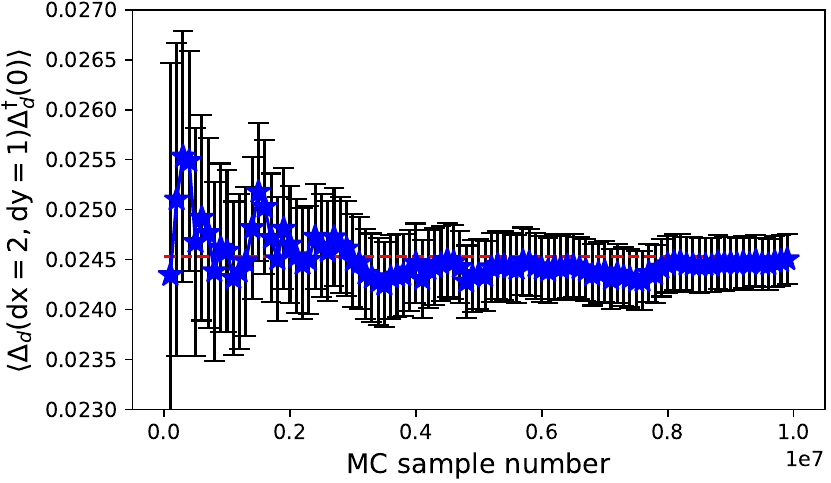}
\caption{Demonstration on the convergence of the pair correlation with respect to MC sample number, for the case of $N_\uparrow=N_\downarrow=5$ and $U=4$ on the $4\times 4$ lattice with PBC, achieved by the $p=1$ wave function. The pair correlation is defined in Eq.(\ref{eq:pair_simple}). }
\label{fig:one_pair_separation_convergence}
\end{figure}
\begin{figure}[t]
\includegraphics[width=\columnwidth]{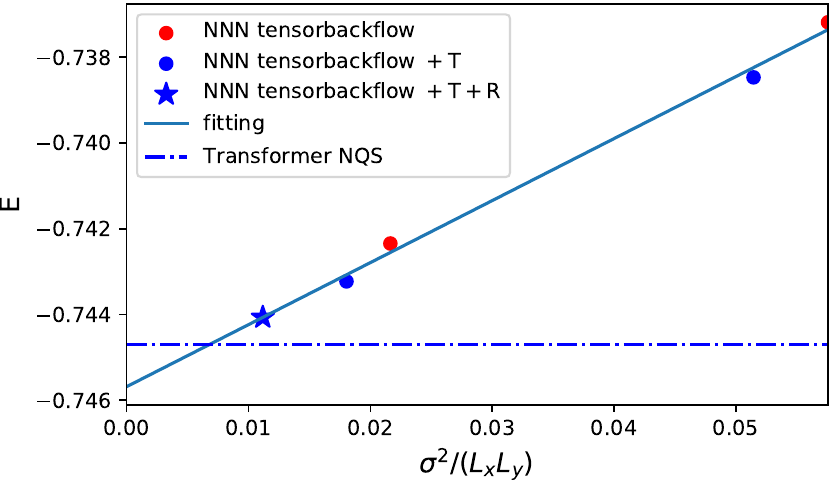}
\caption{Energy extrapolation on the $8\times 8$ lattice. Red and blue dots denote results obtained from tensor-backflow and tensor-backflow along with $T$ projection, respectively. High and low energies depict results with and without one Lanczos step. $T$ and $R$ denote symmetry projections of the half-size translational invariance and the C4 rotational invariance, respectively. Linear fitting is achieved with four solid dots, and the best energy obtained from $T$ and $R$ projection along with a Lanczos step is denoted by the blue star. Although the energy extrapolation is a demonstration of the lowest energy achievable by the method, it shows the precision achieved by Transformer NQS is improvable~\cite{Transformer NQS,symmetry_Hubbard1}. }
\label{fig:8x8_E_extrapolation}
\end{figure}

The Hamiltonian of the $t$-$t'$-$U$ Hubbard model is:
\begin{equation}
        \begin{split}
        \hat{H}=&-t\sum_{\langle ij\rangle,\sigma}(\hat{c}_{i\sigma}^\dagger\hat{c}_{j\sigma}+\mathrm{H.c.})-t'\sum_{\langle\langle ij\rangle\rangle,\sigma}(\hat{c}_{i\sigma}^\dagger\hat{c}_{j\sigma}+\mathrm{H.c.})\\
        &+U\sum_i \hat{n}_{i\uparrow}\hat{n}_{i\downarrow},
        \end{split}
        \label{eq:H}
\end{equation}
where $t$ and $t'$ are strengths for NN and NNN hoppings, respectively. $U$ is the strength of on-site interactions, $\hat{c}_{i\sigma}^\dagger(\hat{c}_{i\sigma})$ creates (destroys) a particle of spin $\sigma$ on the $i$-th site, and the particle number operator $\hat{n}_{i\sigma}=\hat{c}_{i\sigma}^\dagger\hat{c}_{i\sigma}$.
We set $t=1$ through our investigations. %In this work, we consider the range of backflow matching the distance of hopping in the Hamiltonian. Thus, NN and NNN backflows are considered for cases of $t'=0$ and $t'=-0.2t$, respectively.

To benchmark the tensor-backflow wave function, we first compare its pair correlations with exact results for the case of $N_\uparrow=N_\downarrow=5$ on the $4\times 4$ lattice with PBC~\cite{pair_correlation_2D_Hubbard}. The pair correlation is:
\begin{equation}
	\langle \Delta_d(d)\Delta_d^{\dag}(0)\rangle,
	\label{eq:pair_simple}
\end{equation}
with $\Delta_d(d)=c_{d,\uparrow}\sum_\delta f(\delta)c_{d+\delta,\downarrow}$, where the $d$-wave bond-sign structure $f(\pm e_x)=-f(\pm e_y)=1$.
Pair correlation comparisons are depicted in Tab.(\ref{tab:pair_compare}).
In the table, each result of tensor-backflow is obtained by the $p=1$ wave function with $1\times 10^7$ MC sample number.
For each pair separation $\mathbf{d}=(\mathrm{dx}, \mathrm{dy})$, the tensor-backflow result agrees with the exact value.
Based on Eq.(\ref{eq:ws_p1}), the number of determinants scales as $\mathcal{O}(N)$ in a wave function coefficient $w_{p=1}(\mathbf{S})$.
Because the configuration in each determinant is locally changed relative to the original one, only a few matrix elements are changed, thus the efficiency for generating one $w_{p=1}(\mathbf{S})$ is significantly improved by reusing matrix elements.
Because of the high efficiency, the MC sample number for each estimation is as large as $1\times 10^7$, meanwhile the estimation error of Eq.(\ref{eq:pair_simple}) is below $1\times 10^{-3}$.
Fig.(\ref{fig:one_pair_separation_convergence}) demonstrates the estimation convergence for the separation of $(2, 1)$ and $U=4$.

\subsection{$8\times 8$ case}
\begin{table}[]
\begin{tabular}{ccccl}
\hline
\ \ \ \ & \ \ \ \ No Sym\ \ \ \  & \ \ \ \ $T$\ \ \ \  & \ \ \ \ $R$\ \ \ \  & \ \ \ \ $T+R$\ \ \ \  \\ \hline
$E_{p=0}$ & -0.7374(4) & -0.7384(4) & -0.7383(4) & -0.7395(4) \\
$E_{p=1}$ & -0.7424(3) & -0.7432(8) & -0.7431(8) & -0.7441(8)  \\ \hline
\end{tabular}
	\caption{Energy comparisons under different kinds of symmetry projections on the wave function, for the case of $t'=-0.2t$, $U=8$ and $n=0.875$ on the $8\times 8$ lattice with PBC. $T$ and $R$ denote symmetry projections of the half-size translational invariance and the C4 rotational invariance, respectively. Estimation errors are indicated by the last digit in the parentheses.}
	\label{tab:E_8x8}
\end{table}

\begin{figure}[t]
\includegraphics[width=\columnwidth]{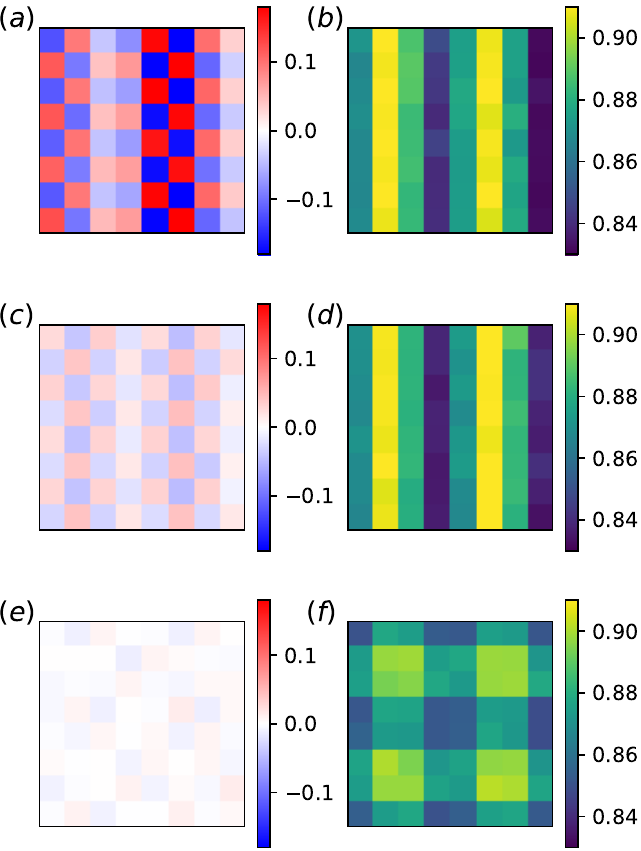}
\caption{Ground state patterns obtained from tensor-backflow without any symmetry projection and with symmetry projections of $T$ and $T+R$, on the $8\times 8$ lattice. Left and right sided figures depict spin density: $s_i^z=(n_i^\uparrow-n_i^\downarrow)/2$ and charge density: $n_i=n_i^\uparrow+n_i^\downarrow$. All results are obtained by $p=1$ wave functions. The tensor-backflow without any symmetry projection(a)(b) with $T$ projection(c)(d) and with $T+R$ projection(e)(f). }
\label{fig:8x8_Sini}
\end{figure}

\begin{figure*}[t]
\includegraphics[width=2\columnwidth]{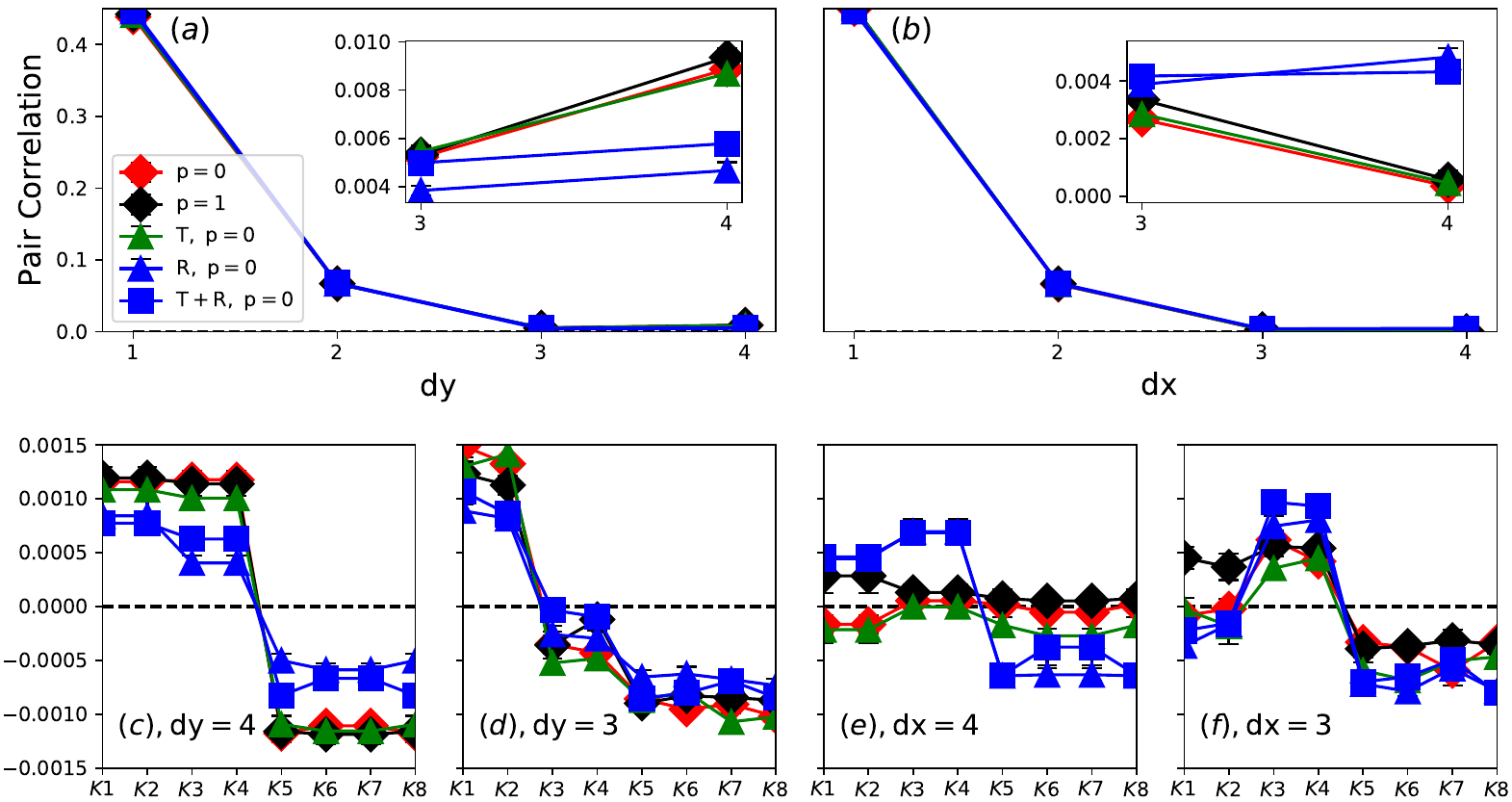}
\caption{$d$-wave pair correlations achieved by different kinds of wave functions on the $8\times 8$ lattice. Definitions of $T$ and $R$ are the same as Fig.(\ref{fig:8x8_E_extrapolation}). $\mathrm{dy}$ and $\mathrm{dx}$ depict pair separations along directions perpendicular to and parallel to the stripe in Fig.(\ref{fig:8x8_Sini}), respectively.
(a,b)The $p$=0 and $p$=1 tensor-backflow states yield similar pair correlations. The $T$ projection produces negligible change relative to the unprojected $p$=0 state. Pair separations near $L/2$ are detailed in insets. The pair correlation is nonzero under the separation $\mathrm{dy}=L/2$, but it decreases to zero under $\mathrm{dx}=L/2$. After the $R$ projection, the pair correlation restores rotational symmetry along both $\mathrm{dx}$ and $\mathrm{dy}$. (c-f)The eight components defined in Eq.(\ref{eq:components}) under pair separations of $L/2$ and $L/2-1$ along x and y axes.}
\label{fig:8x8_SC_order}
\end{figure*}

We first present results for $d$-wave pair correlations at $n=0.875$, $U=8$ and $t'=-0.2t$ on the $8\times 8$ lattice with PBC. The tensor-backflow wave function includes NNN backflow corrections.
The tensor-backflow wave function contains 516096 variational parameters, based on the product of tensor dimensions. The tensor-backflow wave function is initialized from the noninteracting ($U_{UHF}=0$) UHF state.

In this work, we apply several post-optimization improvements to achieve high energy accuracy, we apply projections of half-sized translational symmetry $T$ and C4 rotational symmetry $R$ on the wave function, along with a Lanczos step. Tab.(\ref{tab:E_8x8}) depicts energy comparisons of various posterior optimizations.
From the table, the original tensor-backflow achieves energies of -0.7374 and -0.7424, without and with a Lanczos step, respectively.
The symmetry projection of either $T$ or $R$ improves the energy accuracy, and the energy after a Lanczos step is near -0.743.
Energies are similar for $R$ projections with $\lambda_G=\pm 1$, here we use the $R$ projection with $\lambda_G=1$.
The best energy -0.7441 is achieved with symmetry projections of both $T$ and $R$ along with a Lanczos step.
Compared with the Transformer NQS energy of -0.7447~\cite{Transformer NQS} and the fully symmetrized Pfaffian energy of -0.7506~\cite{symmetry_Hubbard1}, our best energy: -0.7441, is higher by $8.1\times 10^{-4}$ and $8.7\times 10^{-3}$, respectively.

Fig.(\ref{fig:8x8_E_extrapolation}) demonstrates the energy extrapolation for the $8\times 8$ case.
In the figure, $T$ and $R$ denote symmetry projections of the half-size translational invariance and the C4 rotational invariance, respectively.
For both with and without a Lanczos step, energies achieved with symmetry projections are lower than those without symmetry projections.
The energy extrapolation is just a demonstration of the method, and the extrapolated energy at zero variance is estimated as: -0.7455.

Fig.(\ref{fig:8x8_Sini}) depicts ground state patterns of wave functions with different posterior optimizations.
All results in the figure are obtained by $p=1$ wave functions.
The ground state pattern of the tensor-backflow without any symmetry projection is depicted in Fig.(\ref{fig:8x8_Sini})(a)(b), with width-4 stripes obtained successfully. A half-size translational symmetry projection $T$ reduces the spin density meanwhile preserves the charge density, depicted in Fig.(\ref{fig:8x8_Sini})(c)(d). The symmetry projection of both half-size translational invariance $T$ and C4 rotational invariance $R$ reduces the spin density to nearly uniform meanwhile significantly reduces the charge density, depicted in Fig.(\ref{fig:8x8_Sini})(e)(f).
The results of both energy accuracy and ground state patterns demonstrate that symmetry projections can improve the tensor-backflow wave function on the $8\times 8$ lattice, meanwhile a Lanczos step can significantly improve the energy accuracy.

Spin structure factors $S(k_x=\pi, k_y)$ for various kinds of wave functions are depicted in Fig.(\ref{fig:Sk})(a), with the definition: $S(\mathbf{k})=\frac{1}{M^2}\sum_{ij}e^{i\mathbf{k}\cdot (\mathbf{r}_i-\mathbf{r}_j)}s_i^z s_j^z$, where $s_i^z$ is defined in Fig.(\ref{fig:8x8_Sini}).
From the figure, spin structure factors peaks at $k_y=3\pi/4$, indicating the width-4 stripe order. Meanwhile, symmetry projections redistribute the spectral weights thus moderately reduce the peaks.

Based on ground state results, we investigate $d$-wave pair correlations of wave functions with different posterior optimizations on the $8\times 8$ lattice with PBC.
As depicted in Fig.(\ref{fig:8x8_Sini})(a)(b), the tensor-backflow wave function produces the width-4 stripes in the ground state pattern, thus it lacks translational and rotational invariances.
We compare how projections of half-size translational invariance and C4 rotational invariance can affect the $d$-wave pair correlations.
The $s/d$-wave pair-field operator on site $\mathbf{r}_i$ is~\cite{NQS_Pfaffian,SGD_sign2}:
\begin{equation}
	\Delta_{s,d}(\mathbf{r}_i)=\frac{1}{2}[A_{e_x}(\mathbf{r_i})\pm A_{e_y}(\mathbf{r_i})],
	\label{eq:pair-field-operator}
\end{equation}
where $e_{x,y}$ is the unit vector along x/y-axis and $A_\delta(\mathbf{r}_i)=\sum_{\pm}c_{\mathbf{r}_i,\uparrow}c_{\mathbf{r}_i\pm\delta,\downarrow}-c_{\mathbf{r}_i,\downarrow}c_{\mathbf{r}_i\pm\delta,\uparrow}$.
The $s/d$-wave pair correlation is defined as~\cite{NQS_Pfaffian,SGD_sign2}:
\begin{equation}
	P_{s,d}(\mathbf{r}_i,\mathbf{r}_j)=\langle \Delta_{s,d}^\dag(\mathbf{r}_i)\Delta_{s,d}(\mathbf{r}_j)+\Delta_{s,d}(\mathbf{r}_i)\Delta_{s,d}^\dag(\mathbf{r}_j)\rangle.
	\label{eq:pair-correlation}
\end{equation}
By definition, Eq.(\ref{eq:pair-correlation}) can be expanded to eight terms in total, with four terms in the format of $A_{\delta_1}^\dag A_{\delta_2}$ and four terms in the format of $A_{\delta_1} A_{\delta_2}^\dag$, where $\delta_1=e_x,e_y$ and $\delta_2=e_x,e_y$.
Pair correlations for parallel and perpendicular pairs are with $\delta_1=\delta_2$ and $\delta_1\neq \delta_2$, respectively.

\begin{figure*}[t]
\includegraphics[width=2\columnwidth]{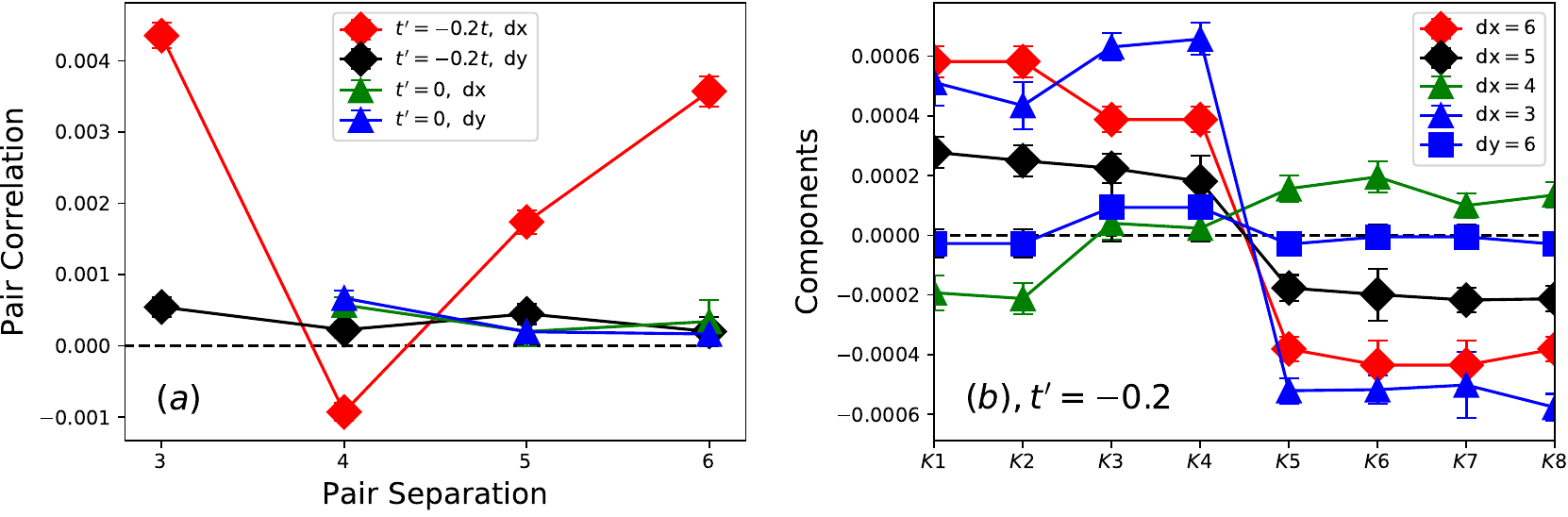}
\caption{For cases of $n=0.875$, $U=8$, $t'=-0.2t$ and $t'=0$ on the $12\times 12$ lattice with PBC. $\mathrm{dx}$ and $\mathrm{dy}$ depict pair separations along directions parallel and perpendicular to the stripe, respectively. Under various pair separations, $d$-wave pair correlations (a) and eight components of the $d$-wave pair correlation (b). All results are obtained within $p=1$ wave functions, and the MC sample number for each estimation is $4.6\times 10^5$. }
\label{fig:12x12_SC_order}
\end{figure*}

In this work, we consider the translational invariant definition of the $d$-wave pair correlation, where the correlation under a pair separation $\mathbf{d}$ is averaged on the whole lattice:
\begin{equation}
	P_d(\mathbf{d})=\frac{1}{M}\sum_{i\in M}P_d(\mathbf{r}_i,\mathbf{r}_i+\mathbf{d}),
	\label{eq:site-averaged-pair-correlation}
\end{equation}
where $\mathbf{d}$ is the pair separation and $M$ is the total site number.
To reveal the bond-sign structure of $d$-wave pair correlation, we consider parallel and perpendicular pairs in the $d$-wave pair correlation defined in Eq.(\ref{eq:site-averaged-pair-correlation}):
\begin{equation}
	\begin{split}
	&C1(\delta_1,\delta_2,\mathbf{d})=\frac{1}{4M}\sum_{i\in M}A_{\delta_1}^{\dag}(\mathbf{r}_i)A_{\delta_2}(\mathbf{r}_i+\mathbf{d}),\\
	&C2(\delta_1,\delta_2,\mathbf{d})=\frac{1}{4M}\sum_{i\in M}A_{\delta_1}(\mathbf{r}_i)A_{\delta_2}^{\dag}(\mathbf{r}_i+\mathbf{d}),
	\end{split}
	\label{eq:eight-components}
\end{equation}
and the eight components from $K1$ to $K8$ are defined:
\begin{equation}
\begin{split}
&K1/2=C1/2(e_x,e_x),\ K3/4=C1/2(e_y,e_y),\\
&K5/6=C1/2(e_x,e_y),\ K7/8=C1/2(e_y,e_x),
\end{split}
\label{eq:components}
\end{equation}
where components of 1, 3, 5, 7 have the form of $A^\dag A$ and components of 2, 4, 6, 8 have the form of $A A^\dag$.
Bonds in components of 1, 2, 3, 4 are parallel to each other and those in components of 5, 6, 7, 8 are perpendicular to each other.
The $d$-wave pair correlation defined in Eq.(\ref{eq:site-averaged-pair-correlation}) is obtainable through the eight components: $\sum_i f(i)Ki$, where $f(i)=\pm 1$ for the first four and the last four components, respectively.

On the $8\times 8$ lattice, we compare components in the $d$-wave pair correlation for various kinds of wave functions in Fig.(\ref{fig:8x8_SC_order}).
Each estimation in the figure is achieved with the MC sample number in the magnitude of $4.6\times 10^{5}$.
In the figure, we consider tensor-backflow wave functions without and with a Lanczos step, half-size translational symmetry projection $T$, C4 rotational symmetry projection $R$ and symmetry projections of both $T$ and $R$. In the figure, all wave functions within symmetry projections are without a Lanczos step.
$\mathrm{dx}$ and $\mathrm{dy}$ depict the pair separation in Eq.(\ref{eq:site-averaged-pair-correlation}) and Eq.(\ref{eq:eight-components}) are along the x-axis and y-axis, respectively.

$d$-wave pair correlations for pair separations along $\mathrm{dy}$ and $\mathrm{dx}$ are depicted in Fig.(\ref{fig:8x8_SC_order})(a)(b).
From the insets, based on both $p=0$ and $p=1$ wave functions,
the tensor-backflow state exhibits a nonzero long-distance pair correlation along $\mathrm{dy}$, whereas the correlation decreases toward zero along $\mathrm{dx}$, reflecting the broken rotational symmetry of the stripe state in Fig.(\ref{fig:8x8_Sini})(a)(b).
$\mathrm{dy}$ and $\mathrm{dx}$ depict pair separations along directions perpendicular to and parallel to the stripe in Fig.(\ref{fig:8x8_Sini}), respectively.
From the insets, the projection of half-size translational invariance doesn't change the $d$-wave pair correlations.
Meanwhile, the C4 projection approximately halves the correlation along $\mathrm{dy}$ and generates an equivalent correlation along $\mathrm{dx}$.
Along both $\mathrm{dy}$ and $\mathrm{dx}$, the projection of $T+R$ has similar results to the projection of $R$.

\begin{table*}[]
\begin{tabular}{cccccc}
\hline
            & \ \ \ \ \ \ \ \ PF1\ \ \ \ \ \ \ \  & \ \ \ \ \ \ \ \ PF2\ \ \ \ \ \ \ \  & \ \ \ \ \ \ \ \ PF2\ \ \ \ \ \ \ \  & \ \ \ \ \ \ \ \ PF3\ \ \ \ \ \ \ \  & \ \ \ \ \ \ \ \ PF3\ \ \ \ \ \ \ \  \\ \hline
$U_{UHF}$   & 4         & 4         & 0         & 4         & 4         \\
Step Number &  2484     &  2152     &  2020     &  1925     &  9013     \\
$E_{p=0}$   & -0.7342(2) & -0.7351(2) & -0.7349(7) & -0.7351(7) & -0.7360(7) \\
$E_{p=1}$   & -0.7396(1) & -0.7400(1) & -0.7400(1) & -0.7401(2) & -0.7408(1)  \\ \hline
\end{tabular}
	\caption{Energy results for the case of $n=0.875$, $U=8$, $t'=-0.2t$ on the $16\times 16$ lattice with PBC, under various pinning fields (PF) and initial $U_{UHF}$. $U_{UHF}$ is the interaction strength in the initial UHF. Step number depicts the energy optimization step number after the pinning field is removed. Estimation errors are indicated by the last digit in the parentheses. }
	\label{tab:E_16x16}
\end{table*}

Fig.(\ref{fig:8x8_SC_order})(c)(d)(e)(f) depict the eight components defined in Eq.(\ref{eq:components}), for various kinds of wave functions and pair separations on the $8\times 8$ lattice.
For the pair separation of $\mathrm{dy}=L/2$, where $L$ is the lattice side length, the $d$-wave bond-sign structure is clearly shown in Fig.(\ref{fig:8x8_SC_order})(c).
From the tensor-backflow wave function for both $p=0$ and $p=1$, because of the separation of $L/2$ on PBC, the pair components satisfy the half-lattice translation relations: $K1=K2$, $K3=K4$, $K5=K8$ and $K6=K7$.
Meanwhile, the $d$-wave bond-sign structure is shown by $K1\approx-K5$,  $K2\approx-K6$, etc. That is, horizontal and vertical bond components have opposite signs.
For the pair separation of $\mathrm{dy}=L/2-1$ depicted in Fig.(\ref{fig:8x8_SC_order})(d), despite the lack of half-size translational invariance for pair components, nonzero $d$-wave pair correlation is also obvious.
Fig.(\ref{fig:8x8_SC_order})(e) depicts the pair separation of $\mathrm{dx}=L/2$.
From both $p=0$ and $p=1$ tensor-backflow wave functions, all eight components are close to zero when $\mathrm{dx}=L/2$.
Fig.(\ref{fig:8x8_SC_order})(f) depicts the pair separation of $\mathrm{dx}=L/2-1$. Local nonzero $d$-wave pair correlation exists based on both $p=0$ and $p=1$ tensor-backflow wave function. Meanwhile the projection of $R$ symmetrizes the pair correlation along $\mathrm{dx}$ and $\mathrm{dy}$.
Note that within the $R$ projection, a component under the same separation length are different along $\mathrm{dx}$ and $\mathrm{dy}$. For example, $K1$ for $\mathrm{dx}=3$ corresponds to $K3$ for $\mathrm{dy}=3$ after the C4 rotational symmetry.

\begin{figure}[t]
\includegraphics[width=1\columnwidth]{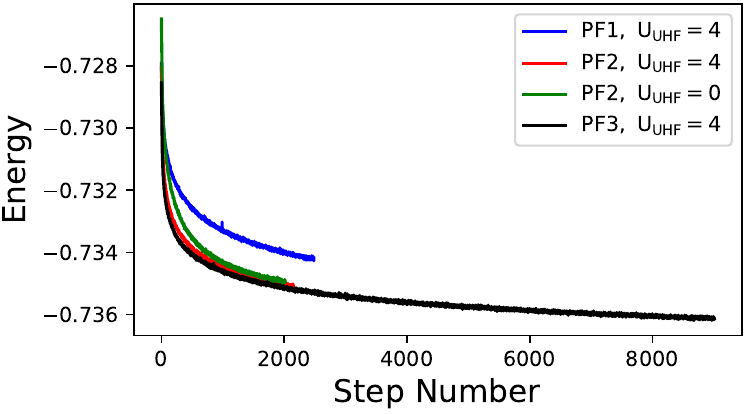}
\caption{Energy convergences for cases of $n=0.875$, $U=8$, $t'=-0.2t$ on the $16\times 16$ lattice with PBC, under various pinning fields (PF) and initial $U_{UHF}$. Step number in the horizontal axis depicts the energy optimization step number after the pinning field is removed. }
\label{fig:16x16_E_convergence}
\end{figure}

\subsection{$12\times 12$, $16\times 16$ case}
For the case of $n=0.875$, $U=8$, $t'=-0.2t$ on the $12\times 12$ lattice with PBC.
The tensor-backflow wave function used in this work was optimized in the previous literature~\cite{TBF}. The wave function was initialized with $U_{UHF}=4$, and the converged energies were -0.7370(3) for $p=0$ and -0.7419(2) for $p=1$.
In contrast to the symmetry-preserving NQS result -0.7443~\cite{NQS byteadance}, the relative energy difference of $p=1$ is $3.2\times 10^{-3}$.
On the $12\times 12$ lattice, energies of $t'=-0.2t$ are -0.7369(5) and -0.7370(5) after the $R$ projection with $\lambda_G=\pm 1$, respectively.
Furthermore, for the $12\times 12$ lattice, the weight of a 90-degree rotated configuration $|w_{p=1}(\hat{T}\mathbf{S})|$ is much smaller than the original one $|w_{p=1}(\mathbf{S})|$, thus configurations around $|\hat{T}\mathbf{S}\rangle$ are neglected when the MC sampling starts from configurations near $|\mathbf{S}\rangle$.
Spin structure factors $S(k_x=\pi,k_y)$ for the $12\times 12$ lattice are depicted in Fig.(\ref{fig:Sk})(b). Because of the side length 12, $S(\mathbf{k})$ doesn't exactly peak at $k_y=3\pi/4$, however it peaks at nearest allowed momentums $k_y=2\pi/3, 5\pi/6$.

We investigate the $d$-wave pair correlations for cases of $t'=-0.2t$ and $t'=0$ on the $12\times 12$ lattice under PBC within the symmetry broken but $p=1$ wave functions. Fig.(\ref{fig:12x12_SC_order}) depicts the pair correlations and components under various pair separations for $t'=-0.2t$ and $t'=0$. Each estimation in the figure is obtained by the $p=1$ wave function within $4.6\times 10^5$ MC samples.
Here $\mathrm{dx}$ and $\mathrm{dy}$ denote pair separations along the x and y directions, respectively.
In Fig.(\ref{fig:12x12_SC_order})(a), $d$-wave pair correlations along $\mathrm{dx}$ and $\mathrm{dy}$ are different because of the symmetry broken state.
For $t'=-0.2t$, nonzero $d$-wave pair correlation is generated along $\mathrm{dx}$.
However, pair correlations for $t'=0$ are close to zero along both $\mathrm{dx}$ and $\mathrm{dy}$. This provides evidence that negative  $t'$ enhances the $d$-wave pair correlations.
Fig.(\ref{fig:12x12_SC_order})(b) depicts the eight components defined in Eq.(\ref{eq:components}), for $t'=-0.2t$.
From the figure, the nonzero pair correlation is accompanied by the expected $d$-wave bond-sign structure at the largest pair separation along $\mathrm{dx}$. Meanwhile, the bond sign structrue varies at other pair separations. Along $\mathrm{dy}$, where zero pair correlation is expected from Fig.(\ref{fig:12x12_SC_order})(a), all components are close to zero.

\begin{figure*}[t]
\includegraphics[width=2\columnwidth]{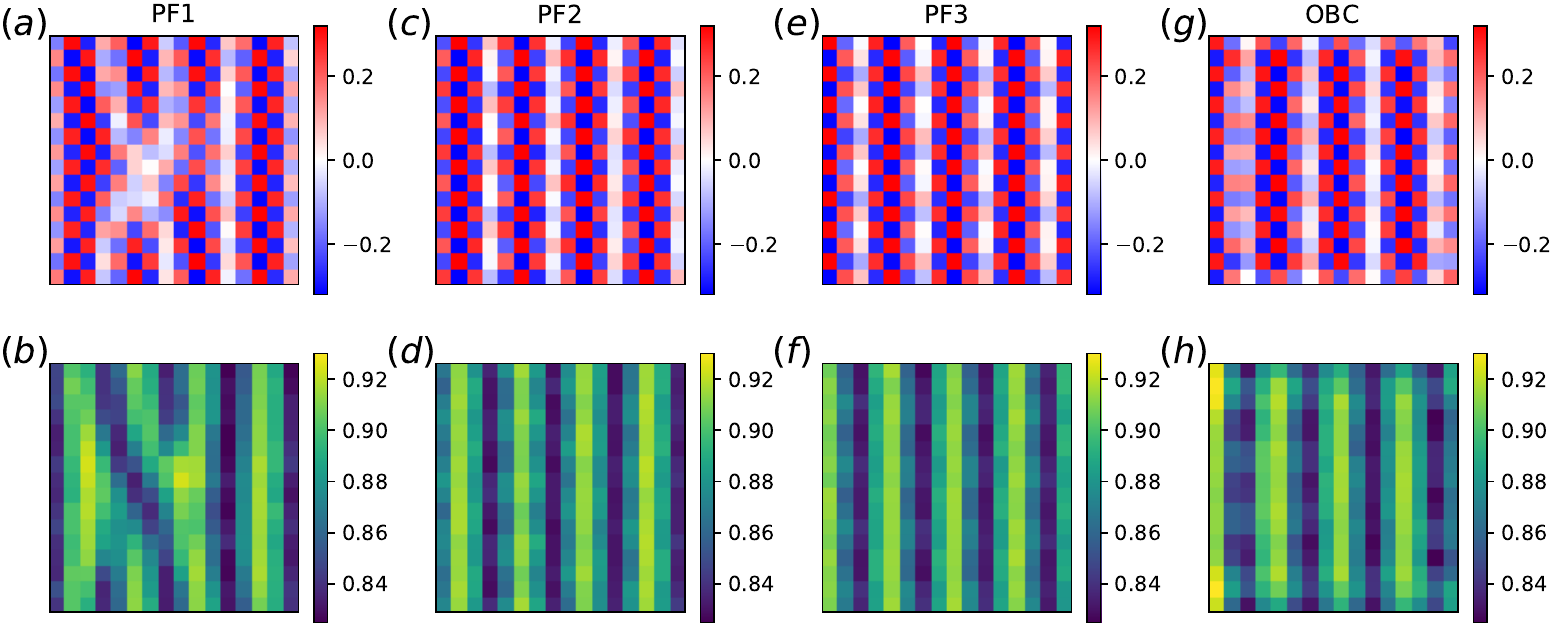}
\caption{Ground state patterns for cases of $n=0.875$, $U=8$, $t'=-0.2t$ on the $16\times 16$ lattice, under various pinning fields (PF) and boundary conditions. Each result is obtained by the $p=1$ wave function with the MC sample number of $10^5$. Under PBC, results from PF1, PF2 and PF3 are depicted by (a)(b), (c)(d) and (e)(f), respectively. Results of PF3 are obtained from the energy of $E_{p=1}=-0.7408$. For results of OBC, the optimization starts from the wave function converged for PF3 under PBC, and the converged energy is $E_{p=1}=-0.7097(1)$.  }
\label{fig:16x16_Sini}
\end{figure*}

We continue investigating cases of $n=0.875$, $U=8$, $t'=-0.2t$ on the $16\times 16$ lattice with PBC.
Unlike the $8\times 8$ and $12\times 12$ lattices, the tensor-backflow wave function requires additional pinning fields to guide the optimization.
We find that the optimization is prone to becoming trapped in local minima even with various $U_{UHF}$.
To achieve the best energy, we compare energies under various setups of pinning fields. The pinning fields are applied by:
\begin{equation}
	\hat{H}_{PF}=\hat{H}+\sum_{i\in PF}h_i(n_i^\uparrow-n_i^\downarrow),
	\label{eq:PF}
\end{equation}
where $\hat{H}$ is the original Hamiltonian defined in Eq.(\ref{eq:H}) and $h_i$ defines the pinning fields.
In this work we consider three kinds of pinning fields, denoted as PF1, PF2 and PF3.
For PF1, $h_i=0.25\times (-1)^{i_x}$ for $i_y=2$ and $h_i=0.25\times (-1)^{i_x+1}$ for $i_y=14$.
For PF2, $h_i=0.25\times (-1)^{i_x}$ for $i_y=2, 10$ and $h_i=0.25\times (-1)^{i_x+1}$ for $i_y=6, 14$.
For PF3, $h_i=0.25\times (-1)^{i_x}$ for $i_y=1, 9$ and $h_i=0.25\times (-1)^{i_x+1}$ for $i_y=5, 13$.
PF2 and PF3 differ only by a translation along $\mathrm{dy}$.
After the UHF is optimized, the pinning fields are applied during the optimization steps, where the MC sample each step is $\mathcal{O}(4.5\times 10^4)$. After several thousands of optimization steps, the ground state pattern according to the pinning fields is established.
Then, the pinning fields are removed and optimizations are continued with approximately $4.5\times 10^5$ MC samples used at each optimization step.

Tab.(\ref{tab:E_16x16}) denotes how converged energies are sensitive to pinning fields, $U_{UHF}$ and optimization step number.
In the table, step number denotes optimization steps used for converging the energy after the pinning fields are removed.
Energy convergences after removing the pinning fields are depicted in Fig.(\ref{fig:16x16_E_convergence}).
From the Tab.(\ref{tab:E_16x16}) and Fig.(\ref{fig:16x16_E_convergence}), PF1 leads to the highest energy. Both PF2 and PF3 give similar energies after a similar step number, regardless of the $U_{UHF}$.
Furthermore, the energy accuracy can be further improved by increasing the step number. As depicted in the figure, for the case of PF3 and $U_{UHF}=4$, increasing the step number from 1925 to 9013 further decreases the $E_{p=1}$ from -0.7401(2) to -0.7408(1), meanwhile, the energy of $E_{p=0}=-0.7360(7)$ corresponds to the Neural Pfaffians~\cite{NQS_Pfaffian}.
Therefore, the energy can be further lowered by continuing the optimization. This observation is consistent with the large numbers of optimization steps employed in recent NQS calculations~\cite{Transformer NQS,NQS byteadance}.

Fig.(\ref{fig:16x16_Sini}) depicts ground state patterns achieved under various kinds of pinning fields. Each spin density and charge density in the figure is obtained by $p=1$ wave functions.
In the figure, results PF1(a)(b) and PF2(c)(d) are obtained by the wave function within the step number about 2000.
Meanwhile, the result of PF3(e)(f) is obtained by the wave function within the step number 9013.
However, ground state patterns of the step number of 1925 and 9013 are similar for PF3.
From the figure, even with the PF1, a width-4 stripe order is in the development.
For PF2 and PF3, width-4 stripe order is successfully established.
We also transfer the wave function converged under PF3 to OBC(g)(h), and the converged energy is $E_{p=1}=-0.7097$. From the figure, width-4 stripe order is preserved in OBC, and the ground state is adjusted according to the boundary condition.
Spin structure factors $S(k_x=\pi,k_y)$ for the $16\times 16$ lattice are depicted in Fig.(\ref{fig:Sk})(c). From the figure, all cases of PF1, PF2, PF3 and OBC have the peak at $k_y=3\pi/4$, indicating the width-4 stripe order.

\begin{figure*}[t]
\includegraphics[width=2\columnwidth]{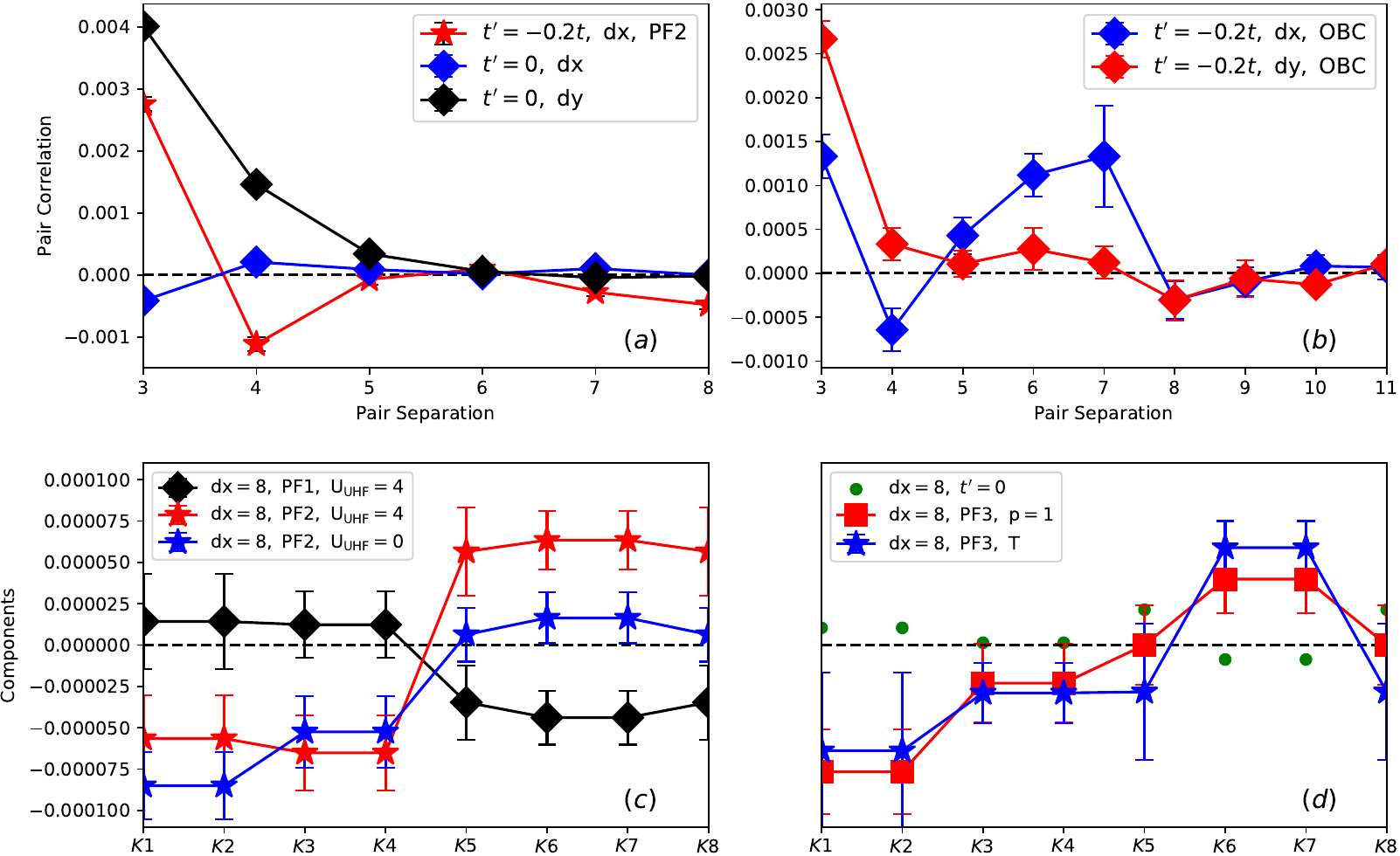}
\caption{$d$-wave pair correlations achieved by different kinds of wave functions on the $16\times 16$ lattice. $\mathrm{dy}$ and $\mathrm{dx}$ depict pair separations along directions perpendicular to and parallel to the stripe in Fig.(\ref{fig:16x16_Sini}), respectively.
All results are obtained using $p=1$ wave functions, except for the PF3 state with $T$ projection. (a)$d$-wave pair correlations under various pair separations, for $t'=-0.2$ and $t'=0$. (b)$d$-wave pair correlations along $\mathrm{dx}$ and $\mathrm{dy}$ for OBC. (c)(d)Components of pair correlations under the largest pair separation along $\mathrm{dx}$. Results of $t'=0$ are denoted by green dots without errorbars for visibility. }
\label{fig:16x16_SC_order}
\end{figure*}

Fig.(\ref{fig:16x16_SC_order}) depicts the $d$-wave pair correlations and components for the $16\times 16$ lattice. In the figure, all results are obtained by $p=1$ wave functions except the case of PF3 with $T$ projection.
Comparisons between $t'=-0.2t$ and $t'=0$ are depicted in Fig.(\ref{fig:16x16_SC_order})(a). From the figure, for the case of $t'=0$, pair correlations decrease to zero in both $\mathrm{dx}$ and $\mathrm{dy}$, thus we find no evidence for robust long-distance $d$-wave pairing correlations at this lattice size.
For the case of $t'=-0.2t$, nonzero pair correlations exist under the largest pair separation along $\mathrm{dx}$.
Components defined in Eq.(\ref{eq:components}) for $\mathrm{dx}=L/2$ are depicted in Fig.(\ref{fig:16x16_SC_order})(c)(d).
From Fig.(\ref{fig:16x16_SC_order})(c), $d$-wave bond-sign structure are generated for both PF1 and PF2. Although components of PF1 and PF2 have the opposite global sign.
%This is normal for the wave function without explicit symmetry enforcements.
Fig.(\ref{fig:16x16_SC_order})(d) depicts results of PF3 and compares the components to the case of $t'=0$.
Because pinning fields of PF2 and PF3 differs only by a translation along $\mathrm{dy}$, $d$-wave components from PF2 and PF3 are similar. For comparisons, components for $t'=0$ are denoted in Fig.(\ref{fig:16x16_SC_order})(d) without errorbars for visibility. From the figure, all components of $t'=0$ are close to zero, matching the zero $d$-wave pair correlations in Fig.(\ref{fig:16x16_SC_order})(a).

The pair components in Eq.(\ref{eq:components}) are difficult to converge for the $16\times 16$ lattice, as they require a large MC number. Convergence of pair components are depicted in Fig.(\ref{fig:pair_correlation_convergence}).
In the figure, under the pair separation $\mathrm{dx}=L/2$, each estimation is obtained by the $p=1$ wave-function, 829440 and 645120 MC samples are used for $t'=-0.2t$ and $t'=0$, respectively.
From the figure, $d$-wave bond-sign structure is revealed for the $t'=-0.2t$ case. However, for $t'=0$, all components are close to zero.

We study the case of OBC to investigate $d$-wave pair correlations under larger pair separations.
To preserve the width-4 stripe order, the wave function is transferred from the case of PF3 under PBC, and it has been optimized for 1324 steps. The final converged energies are $E_{p=0}=-0.7048(7)$ and $E_{p=1}=-0.7097(1)$.
For OBC, instead of averaging $d$-wave pair correlations on the whole lattice denoted in Eq.(\ref{eq:site-averaged-pair-correlation}), there are $i_x\in[3,14], i_y=3$ for $\mathrm{dy}\in [3, 11]$ and $i_x=3, i_y\in[3,14]$ for $\mathrm{dx}\in [3, 11]$.
From Fig.(\ref{fig:16x16_Sini})(g)(h) and Fig.(\ref{fig:Sk})(c), the width-4 stripe order is preserved under OBC.
Fig.(\ref{fig:16x16_SC_order})(b) depicts $d$-wave pair correlations for OBC.
Different from PBC, pair correlations are close to zero under largest pair separations along both $\mathrm{dx}$ and $\mathrm{dy}$ for OBC.

\subsection{Pair structure factor and pair-density matrix eigenvalues}
Beyond the real-space bond-sign analysis, we investigate whether the tensor-backflow states exhibit off-diagonal long-range order (ODLRO)~\cite{ODLRO} by examining the pair structure factor and the eigenvalue spectrum of the pair-density matrix on $8\times 8$, $12\times 12$ and $16\times 16$ lattices.

To determine the type of the ground state, we calculate the pair structure factor~\cite{pair structure factor}:
\begin{equation}
	S_d(\mathbf{k})=\frac{1}{M^2}\sum_{ij}e^{i\mathbf{k}\cdot (\mathbf{r}_i-\mathbf{r}_j)}\langle\Delta_d^\dag(\mathbf{r}_i)\Delta_d(\mathbf{r}_j) \rangle,
	\label{eq:S_d}
\end{equation}
and eigenvalues of the pair-density matrix~\cite{eigenvalues}:
\begin{equation}
	P_{ij}=\langle \Delta_d^\dag(\mathbf{r}_i)\Delta_d(\mathbf{r}_j)\rangle,
	\label{eq:pair_density_matrix}
\end{equation}
where $\Delta_d$ is defined in Eq.(\ref{eq:pair-field-operator}).
We verify on $8\times8$ that the Lanczos step substantially improves the energy while leaving the pair correlations nearly unchanged. This motivates evaluating the computationally more demanding pair-density matrix using $p=0$ wave functions on the larger lattices.
To reduce the computational cost, we calculate the pair structure factor by $p=0$ wave functions, each matrix element $\langle\Delta_d^\dag \Delta_d\rangle$ is estimated using 115200 MC samples.

\begin{figure}[t]
\includegraphics[width=\columnwidth]{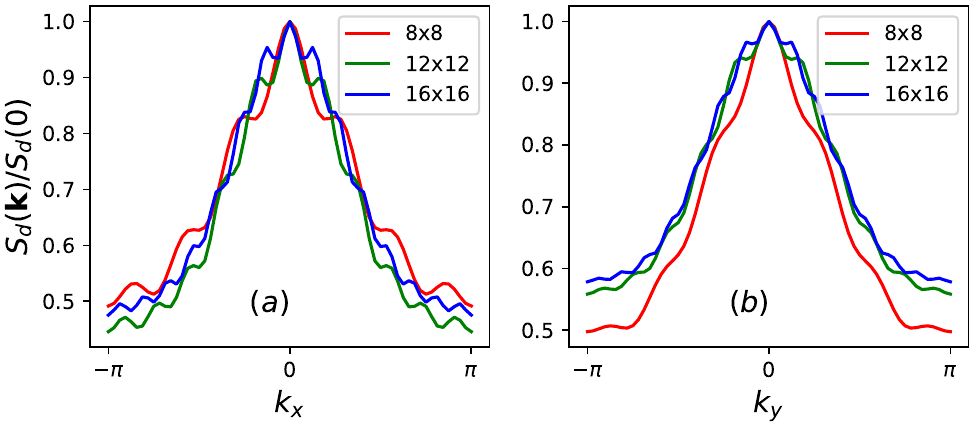}
\caption{Pair structure factors for $k_y=0$(a) and $k_x=0$(b) on $8\times 8$, $12\times 12$ and $16\times 16$ lattices, each result is obtained by the $p=0$ wave function. }
\label{fig:Sk_pair}
\end{figure}
\begin{figure}[t]
\includegraphics[width=\columnwidth]{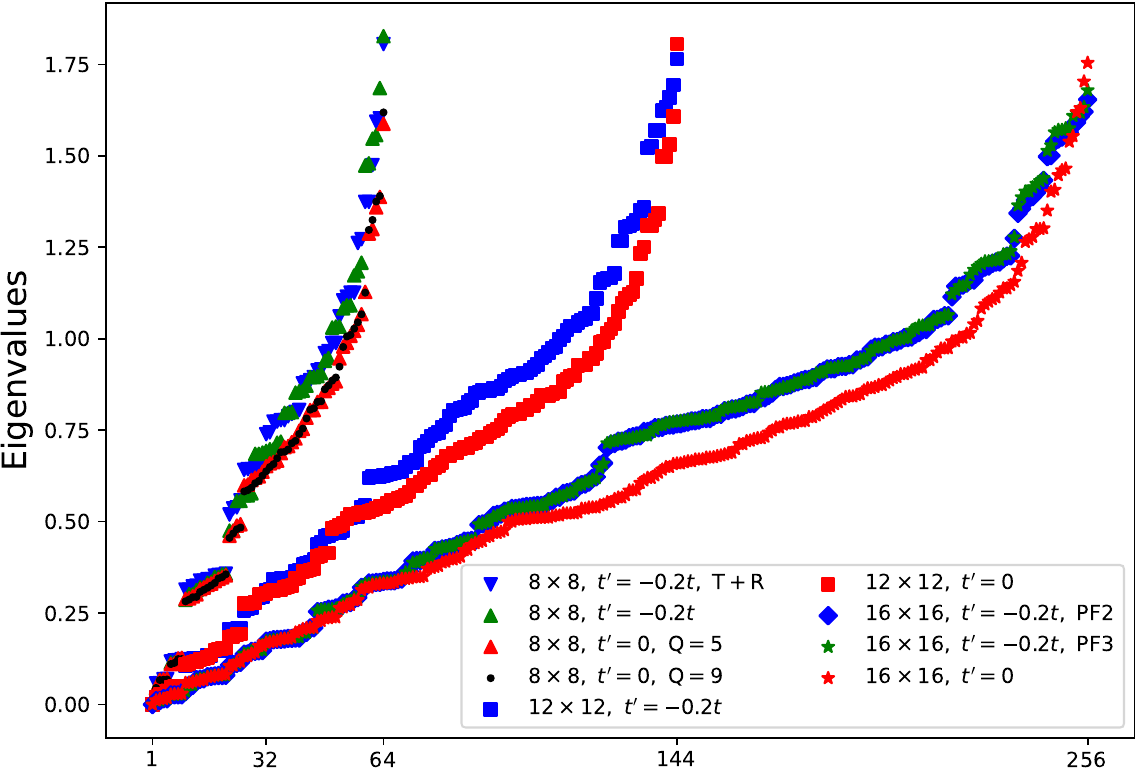}
\caption{Eigenvalue spectra of the $d$-wave pair-density matrices for the $8\times 8$, $12\times 12$ and $16\times 16$ lattices. Each result is obtained by the $p=0$ wave function. }
\label{fig:eigenvalues}
\end{figure}

Fig.(\ref{fig:Sk_pair})(a)(b) depict pair structure factors for $k_y=0$ and $k_x=0$, respectively.
From the figure, pair structure factors are centered at $\mathbf{k}=0$ for all lattice sizes, indicating that the dominant pairing correlations occur at zero center-of-mass momentum.
However, the peak does not visibly sharpen with increasing lattice size over the sizes studied, consistent with predominantly short-range pairing correlations.

Besides the pair structure factor, Fig.(\ref{fig:eigenvalues}) depicts eigenvalues of pair-density matrices for $t'=-0.2t$ and $t'=0$ on $8\times 8$, $12\times 12$ and $16\times 16$ lattices.
From the figure, on the $8\times 8$ lattice, the leading $d$-wave pair-density matrix eigenvalue is enhanced for $t'=-0.2t$ relative to $t'=0$.
For the $8\times 8$ lattice, although symmetry projections improve the variational energy effectively, largest eigenvalues with and without symmetry projections are similar.
For the $8\times8$ lattice, we evaluate the $t'=0$ state using NN backflow ($Q$=5) and NNN backflow ($Q$=9) and the resulting spectra are similar, indicating that the enhanced pairing spectrum at $t'=-0.2t$ is not simply caused by the larger backflow range.

On larger lattices such as $12\times 12$ and $16\times 16$, the largest eigenvalue of $t'=-0.2t$ is not higher than that of $t'=0$.
However, near the largest eigenvalue of $t'=-0.2t$, the $d$-wave pair spectral weight is distributed among several nearly degenerate leading eigenmodes rather than being concentrated as strongly in a single dominant eigenmode, suggesting a larger number of nearly degenerate leading $d$-wave pair modes for $t'=-0.2t$.
Similar to the $8\times 8$ case, the leading part of the pair-density matrix spectrum is generally shifted toward larger eigenvalues for $t'=-0.2t$, indicating stronger $d$-wave pair correlations distributed among several modes.

In Fig.(\ref{fig:eigenvalues}), each matrix element is estimated using 115200 MC samples. The resulting spectra show consistent qualitative trends among independently optimized wave functions. However, we do not use small differences between individual nearly degenerate eigenvalues as quantitative evidence.
ODLRO requires the largest eigenvalue of the pair-density matrix to scale extensively, $\lambda_{max}\propto M$ in the thermodynamic limit~\cite{ODLRO,eigenvalues}.
Based on Fig.(\ref{fig:eigenvalues}), the leading eigenvalue does not show growth proportional to the system size over the lattice sizes studied.
Based on Fig.(\ref{fig:16x16_SC_order})(b), the $d$-wave pair correlation decreases to zero at large pair separations under OBC, thus the nonzero pair correlation under PBC should not be interpreted as robust ODLRO.
Both Fig.(\ref{fig:eigenvalues}) and Fig.(\ref{fig:16x16_SC_order})(b) suggest enhanced $d$-wave pairing fluctuations rather than established superconducting ODLRO.

\section{Discussions and conclusions}
\label{sec:conclusions}

\begin{figure*}[t]
\includegraphics[width=2\columnwidth]{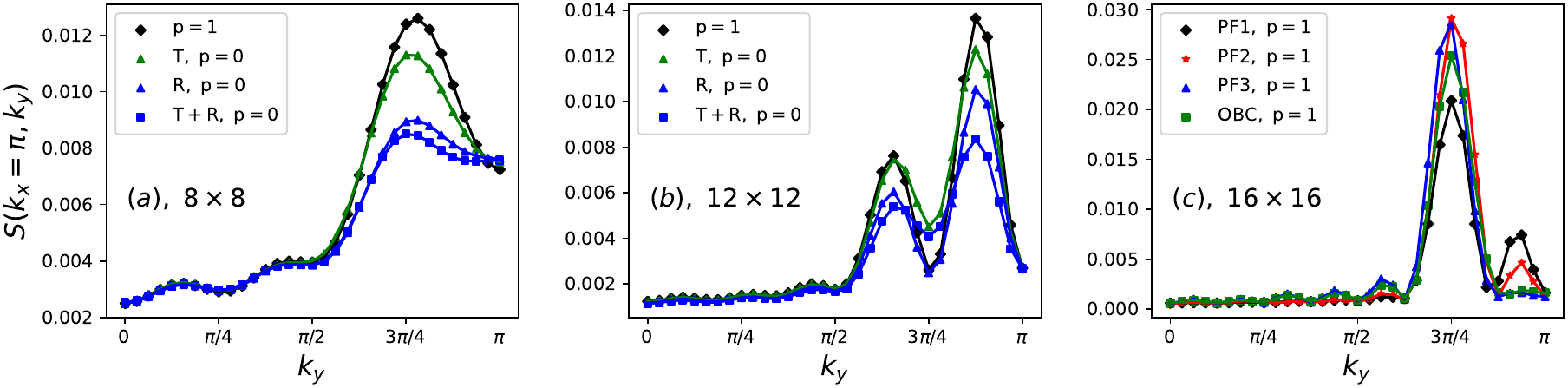}
\caption{Spin structure factors for cases of $n=0.875$, $U=8$, $t'=-0.2t$ on lattices of $8\times 8$(a), $12\times 12$(b) and $16\times 16$(c). Unless otherwise specified, all results are obtained under PBC.}
\label{fig:Sk}
\end{figure*}
	\begin{figure*}[t]
\includegraphics[width=2\columnwidth]{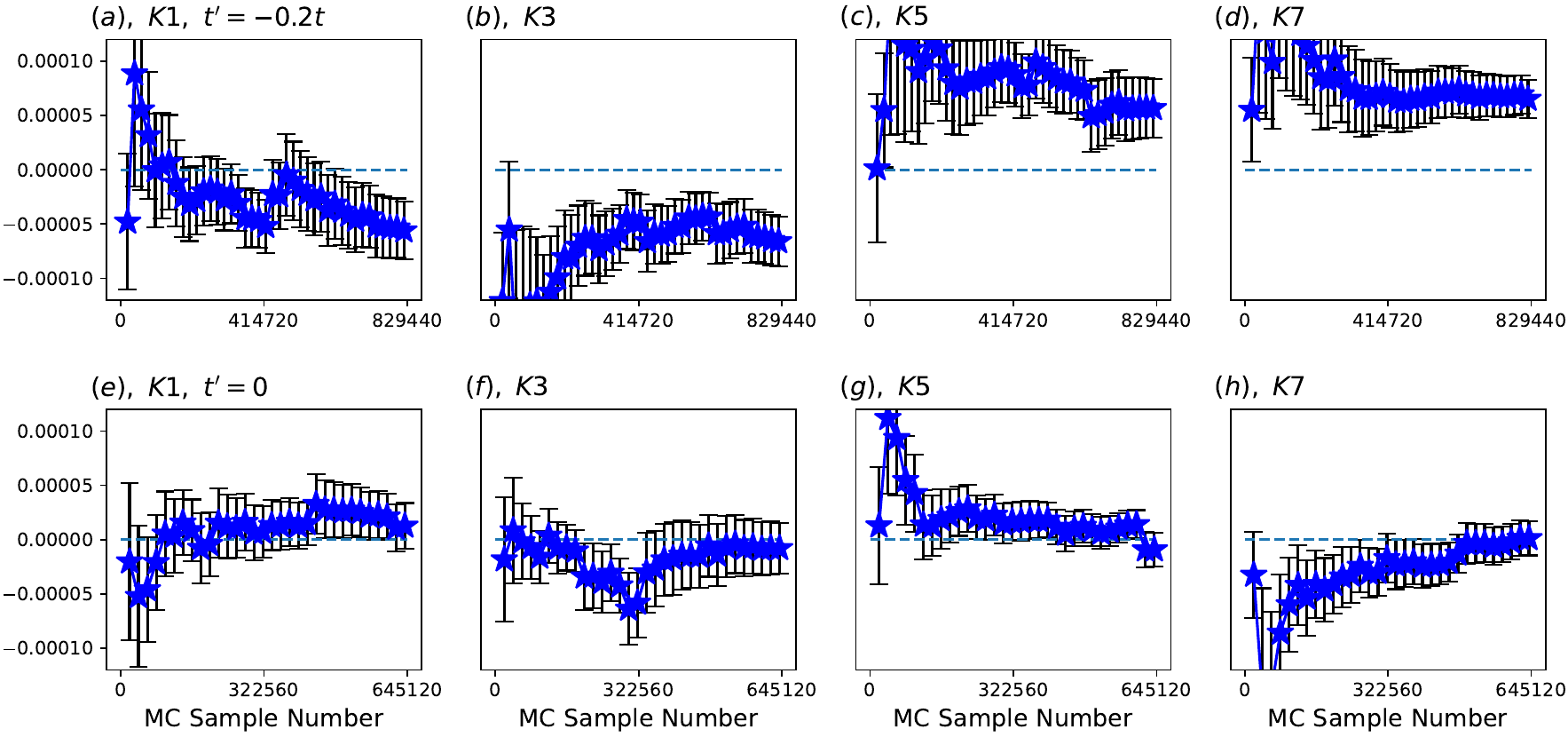}
\caption{Convergence of pair correlation components defined in Eq.(\ref{eq:components}) under $\mathrm{dx}=L/2$ for $t'=-0.2t$ and $t'=0$ on the $16\times 16$ lattice. All results are obtained by $p=1$ wave-functions. PF2 is employed for initializing the $t'=-0.2t$ case.}
\label{fig:pair_correlation_convergence}
\end{figure*}

In this work, we apply the tensor-backflow wave function combined with various posterior optimization methods to investigate the $d$-wave superconductivity in the $t$-$t'$-$U$ Hubbard model on square lattices.
We compare cases of both $t'=-0.2t$ and $t'=0$ on the same square lattice to investigate the effects of $t'=-0.2t$.

We obtain energies competitive with current state-of-the-art NQS results.
On the $8\times 8$ lattice, we use symmetry projections along with a Lanczos step to improve the energy accuracy. The best energy achieved in this work for $t'=-0.2t$ is -0.7441. The relative energy differences are $8.1\times 10^{-4}$ and $8.7\times 10^{-3}$ compared with Transformer NQS~\cite{Transformer NQS} and fully symmetrized Pfaffian~\cite{symmetry_Hubbard1}, respectively.
On the $12\times 12$ lattice, we use the tensor-backflow wave function optimized in the previous literature~\cite{TBF}, within the energy $E_{p=1}=-0.7419(2)$, and the relative energy difference to the symmetry-preserving NQS result is $3.2\times 10^{-3}$.
On the $16\times 16$ lattice, we optimize the wave function using pinning fields.
We compare states obtained with different pinning field protocols at comparable optimization stages. and find that the width-4 stripe order gives the lowest energy, and the relative energy difference is $4.4\times 10^{-3}$ relative to the symmetry-preserving NQS.
Furthermore, we show that the energy accuracy can be improved by simply increasing optimization steps, however the qualitative stripe pattern remains unchanged.

In this work, although a Lanczos step substantially lowers the energy, comparison of the $p=0$ and $p=1$ states shows that the qualitative $d$-wave pairing behavior is already captured by the original tensor-backflow state.
This illustrates that, in a regime containing several nearly degenerate competing states, improvements in variational energy do not necessarily lead to equally large changes in order-sensitive observables. It is therefore important to compare both energies and order-sensitive observables across competing variational states.

We have compared real-space $d$-wave pair correlations, pair structure factors, and pair-density matrix spectra for $t'=-0.2t$ and $t'=0$. On the $8\times 8$ lattice, the leading pair-density matrix eigenvalue is enhanced for $t'=-0.2t$.
On the $12\times 12$ and $16\times 16$ lattices, negative $t'$ enhances the leading portion of the pair-density matrix spectrum, with the spectral weight distributed among several nearly degenerate modes. Because the $p=0$ wave function contains no explicit pairing determinant or pairing function, these enhanced $d$-wave correlations arise from the configuration-dependent backflow structure of the variational state. Overall, negative $t'$ enhances $d$-wave pairing fluctuations in the width-4 striped state, but neither the pair structure factor nor the finite-size scaling of the pair-density matrix spectrum establishes superconducting ODLRO up to $16\times 16$.

\section{Acknowledgement}
X. Liang thanks the inspirational discussion with Daniele Guerci and Shiwei Zhang.
X. Liang discussed with GPT-5.6 Sol and Gemma4 on concepts and definitions in the manuscript preparation. All equations were derived by hand but verified by GPT-5.6 Sol.
The author acknowledges William \& Mary Research Computing and the Texas Advanced Computing Center (TACC) at The University of Texas at Austin for providing computational resources that have contributed to the results reported within this paper.

\begin{appendices}
	\section{Spin structure factors}

	Fig.(\ref{fig:Sk}) depicts spin structure factors $S(k_x=\pi,k_y)$ for cases of $n=0.875$, $U=8$, $t'=-0.2t$ on lattices of $8\times 8$(a), $12\times 12$(b) and $16\times 16$(c). $p=0$ and $p=1$ wave function are used for cases with and without symmetry projections, respectively.

	\section{Pair correlation convergence on $16\times 16$ lattice}

Fig.(\ref{fig:pair_correlation_convergence}) depicts the convergence of pair correlation components defined in Eq.(\ref{eq:components}) for $t'=-0.2t$ and $t'=0$ on the $16\times 16$ lattice. All results are obtained by $p=1$ wave-functions. PF2 is employed for initializing the $t'=-0.2t$ case.
Because of PBC, four components are shown for each $t'$ case.
From the figure, we use a large MC sample number to reveal bond-sign structures for $t'=-0.2t$ and $t'=0$.
For $t'=-0.2t$, $d$-wave pairing bond-sign structure is revealed. However for $t'=0$, all components are close to zero.

\end{appendices}

\end{document}